\documentclass[submission,copyright,creativecommons]{eptcs}
\providecommand{\event}{Gandalf 2011}

\usepackage{epsfig}

\newcommand{\expand}[1]{\mathit{Exp}(H)}
\newcommand{\post}[1]{\mathit{post}(#1)}
\newcommand{\conf}[1]{\mathit{conf}(#1)}
\newcommand{\rev}[1]{\mathit{rev}(#1)}
\newcommand{\postfix}[2]{#1/_{#2}}

\newcommand{\define}{:=}
\newcommand{\implies}{\Longrightarrow}
\newcommand\bijects[2]{\colon #1\leftrightarrow #2}
\newcommand{\C}{\mathcal{C}}
\newcommand{\Cut}{\mathit{Cut}}
\newcommand{\Mark}{\mathit{Mark}}

\usepackage{tabularx}
\usepackage[table]{xcolor}
\usepackage{latexsym}
\usepackage{amssymb}

\usepackage[version=0.96]{pgf}
\usepackage{tikz}
\usepackage{url}
\usetikzlibrary{arrows,shapes,snakes,automata,backgrounds,petri}

\usepackage{algorithm}
\usepackage{algorithmic}
\usepackage{booktabs}
\usepackage[latin1]{inputenc}
\newcommand\by{\begin{eqnarray}}
\newcommand\ey{\end{eqnarray}}
\newcommand\bys{\begin{eqnarray*}}
\newcommand\eys{\end{eqnarray*}}

\newcommand\NN{\makebox[.1em][l]{$\mathsf{I}$}\makebox[.1em][l]{$\mathbf{N}$}\hspace{0.45em}\ }
\newtheorem{definition}{Definition}
\newtheorem{lemma}{Lemma}
\newtheorem{theorem}{Theorem}
\newtheorem{example}{Example}
\newcommand\emogl{\sqsubseteq}
\newcommand\conflict{\mathbin{\#}}

\newcommand\concurrent{\mathbin{\mathbf{co}}}
\newcommand\netn{{\mathit{N}}}

\newcommand\node[1]{\mathit{#1}}
\newcommand\bel{\begin{Lemma}}
\newcommand\enl{\end{Lemma}}
\newcommand\bep{\begin{Proposition}}
\newcommand\enp{\end{Proposition}}
\newcommand\bdf{\begin{definition}}
\newcommand\edf{\end{definition}}
\newcommand\bet{\begin{theorem}}
\newcommand\ent{\end{theorem}}
\newcommand\brem{\begin{rem}}
\newcommand\erem{\end{rem}}
\newcommand\bum{\begin{enumerate}}
\newcommand\eum{\end{enumerate}}
\newcommand\bit{\begin{itemize}}
\newcommand\eit{\end{itemize}}
\newcommand\reach{\mathbf{R}}

\newcommand\bepr{\par {\bf Proof:}\ }
\newcommand\eepr{\nopagebreak \hspace*{1cm} \nolinebreak \hspace{\fill} \nolinebreak $\Box$}

\newcommand\witness{\mathbf{wit}}

\newcommand\cone[1]{\lceil{#1}\rceil}

\newcommand\precone[1]{\lfloor{#1}\rfloor}
\newcommand{\preset}[1]{{{}^\bullet{#1}}}
\newcommand{\postset}[1]{{#1}^\bullet}

\newcommand\flow{\mathit{F}}

\newcommand\otransn{\mathit{e}}

\newcommand\trans{\mathit{T}}

\newcommand\heightsym{\mathcal{H}}
\newcommand\height[1]{\mathcal{H}(#1)}

\newcommand{\conflictsymb}{\#}
\newcommand\flict[1]{\conflictsymb[#1]}

\newcommand\cov[1]{\mathord{\reveals}[#1]}
\newcommand\reveals{\mathbin{\triangleright}}

\newcommand\move[1]{\stackrel{#1}{\longrightarrow }}

\newcommand\omove[1]{\stackrel{#1}{\leadsto }}
\begin{document}
\title{Computing the Reveals Relation in Occurrence Nets}
\author{Stefan Haar
\institute{INRIA and LSV, Ecole Normale Sup\'erieure de Cachan and CNRS \\  61, ave. du Pr\'esident Wilson, 94230 Cachan, France}
\and Christian Kern\thanks{The author was supported by the DFG Graduiertenkolleg 1480 (PUMA).}
\institute{TU M\"unchen, Boltzmannstr. 3, 85748 Garching, Germany}
\and Stefan Schwoon
\institute{LSV, Ecole Normale Sup\'erieure de Cachan and CNRS, and INRIA \\  61, ave. du Pr\'esident Wilson, 94230 Cachan, France}
}
\def\titlerunning{Computing the Reveals Relation in Occurrence Nets}
\def\authorrunning{S. Haar, C. Kern, S. Schwoon}

\maketitle%

\begin{abstract} 
Petri net unfoldings are a useful tool to tackle state-space
explosion in verification and related tasks. Moreover, their structure allows to access \emph{directly} the relations of causal precedence, concurrency, and conflict between events. Here, we explore the data structure further, to determine the following relation: event $a$
is said to \emph{reveal} event $b$ iff the occurrence of a implies that b
inevitably occurs, too, be it before, after, or concurrently with $a$.
Knowledge of \emph{reveals} facilitates in particular the analysis of
partially observable systems, in the context of diagnosis, testing, or
verification; it can also be used to generate more concise representations
of behaviours via abstractions.
The reveals relation was previously introduced in the context of fault
diagnosis, where it was shown that the reveals relation
was decidable: for a given pair $a,b$ in the unfolding $U$ of a safe Petri
net $N$, a finite prefix $P$ of $U$ is sufficient to decide whether or not
$a$ reveals $b$. In this paper, we first considerably improve the bound on
$|P|$. We then show that there exists an efficient algorithm for computing
the relation on a given prefix. We have implemented the algorithm and report
on experiments.
\end{abstract}

\paragraph*{Topics:} 
Structure and behaviour of Petri Nets; partial-order theory of concurrency; automatic analysis

\section{Introduction}

Petri nets (see e.g. \cite{Peterson,Murata}) and their
partial-order unfoldings \cite{McMillan,esparza-1,KhoKouVog:CAV02}
have long been used in model checking. Their crucial feature is
the partial-order representation of concurrency, allowing to escape
from the state-space-explosion problem that is brought about by
the use of interleaving semantics~\cite{cantavoid}.

In this paper, we will focus on the  problem of determining
the following relation: an event $a$ is said to \emph{reveal} another event
$b$ iff, whenever $a$ occurs, the occurrence of $b$ is inevitable.
This does not imply that $a$ and $b$ are causally related (though they may be);
in fact, $b$ may have occurred before $a$, lie in the future of $a$, or
even be concurrent to $a$. To some degree, this relation is complementary to the
well-known \emph{conflict} relation: $a$ and $b$ are in conflict if the
occurrence of $a$ implies that the occurrence of $b$ is impossible. Notice
however that the conflict relation is symmetric while \emph{reveals} is not.

We further emphasize that the \emph{reveals} relation is essentially a non-temporal relation, as opposed  to temporal properties or the \emph{synchronic distance} of e.g. \cite{Goltz87,Reisigbook,zhao:synchronic}. The latter measures the 
quantitative degree of independency in the repeated occurrences of two net transitions, whereas $a \reveals b$ holds if and only if 
event $a$ \emph{implies} event $b$.

The reveals relation was first introduced in \cite{CDC07}; more properties and discussions of its applications are given in  \cite{Haar-tac10}.  An important motivation for studying \emph{reveals} 
lies in the partial observability of many systems  in applications 
such as those related to fault diagnosis.  The idea is that $a\reveals b$
implies that it suffices to observe $a$ to infer occurrence of $b$;
conversely, $b$ does not have to be observable itself, provided $a$
or any other event that reveals $b$ is observable. 

This \emph{binary} relation is the topic of the present article.
Recently, \cite{BalChaHaa11} gave generalizations that include a reveals relation connecting pairs of \emph{sets} of events; however, even in this general setting the binary relation turns out to play a central role. Its exploration and effective computation remains therefore an important task, not only for the structural theory.
In fact, 
$\reveals$ is relevant in general for opacity-related properties and tasks concerning concurrent systems; potential and actual applications include verification diagnosability  (see \cite{Haar-tac10,CDC09}) and other properties, conformance testing, synthesis of controllers and adaptors.

Concerning the task at hand, note that it was shown in 
\cite{Haar-tac10} that the \emph{reveals} relation can
be effectively computed for unfoldings of safe nets. For each pair of events
$(a,b)$, a suitable finite prefix whose height exceeds that of $a$ and $b$ by
at most a uniform bound, is sufficient to verify if $a$ reveals $b$. Here,
we make the following contributions:
\begin{itemize}
\item We considerably improve the bound on the size of the finite prefix
  needed to decide whether $a$ reveals $b$. While the previous bound
  seemed to make this decision impracticable, the new bound gives much
  more hope to determine the relation in practice.
\item Motivated by this, we discuss an efficient algorithm that computes the
  entire reveals relation within a given prefix. The algorithm can be implemented
  completely with bitset operations.
\item We have implemented the algorithm and report on experiments, notably
  on the following questions: how big is the prefix necessary to determine
  the reveals relation, and how much time does it take to compute said
  relation on a given prefix?
  Concerning the second question, the algorithm turns out to be suitably fast;
  it works on prefixes with tens of thousands of events in a few seconds, and
  usually takes less time than the actual construction of the prefix.
\end{itemize}

We proceed as follows: Section~\ref{sec:def} introduces Petri nets,
their unfoldings, the reveals relation, and some of its salient properties.
Section~\ref{sec:reveal} gives the new bound on the size of the prefix.
Section~\ref{sec:alg} presents an algorithm for computing \emph{reveals} on a
given prefix, and Section~\ref{sec:exp} presents the experiments. We conclude in
Section~\ref{sec:conc}.



\section{Definitions}\label{sec:def}

This section introduces central definitions and facts about Petri nets, their unfoldings, and the
reveals relation. While most definitions and some results would be valid 
in the case of Petri nets that are bounded, but not 1-bounded, our main interest is in 1-bounded (aka safe) nets. Moreover, lifting to non-safe nets brings little additional insight
but makes arguments much more technical and cumbersome; we therefore
chose to focus on safe nets.

\subsection{Petri nets}
A \emph{Petri net} is a triple ${\netn=(P,T,F,M_0)}$,
where $P$ and $T$ are disjoint sets of \emph{places} and
\emph{transitions}, respectively, and
$F\subseteq (P\times T)\cup(T\times P)$
is the \emph{flow relation}. Any function $M\colon P\to\NN$ is called a
\emph{marking}, and $M_0$ is the \emph{initial marking}.
By \emph{node}, we shall mean an element from the set $P\cup T$.

In figures (e.g., the left-hand side of Figure~\ref{fig:netandon}), circles represent places, rectangular boxes represent transitions, and directed edges represent $F$. A marking $M$ is represented by black tokens.

For a node $x$, call $\preset{x} := \{\,x' \mid (x',x)\in F\,\}$
the \emph{preset}, and $\postset{x} := \{\,x'  \mid (x,x')\in F\,\}$
the \emph{postset} of~$x$. Moreover, for any set $X\subseteq P\cup T$, set 
\begin{eqnarray*}
\preset{X}\define\bigcup_{x\in X}\preset{x} &and& \postset{X}\define
\bigcup_{x\in X}\postset{x}.
\end{eqnarray*}
Transitions induce a \emph{firing relation}
among markings, as follows: Let $M,M'$ be markings and $t$ a transition.
Then we write $M\move{t}M'$ iff $M(p)\ge1$ for every $p\in\preset{t}$ and
$M'(p)=M(p)-1$ if $p\in\preset{t}\setminus\postset{t}$,
$M'(p)=M(p)+1$ if $p\in\postset{t}\setminus\preset{t}$,
and $M'(p)=M(p)$ otherwise. In words, we also say
that $t$ is \emph{enabled} in $M$, and that \emph{firing} it leads to $M'$.

A finite sequence $\sigma:=t_1\ldots t_k$ of transitions is a \emph{run} iff
$M_0\move{t_1}M_1\cdots\move{t_k}M_k$ for some markings $M_1,\ldots,M_k$;
if such a run exists, then $M_k$ is said to be \emph{reachable}. The
set of reachable markings is denoted $\reach(N)$.
A net is said to be \emph{safe} if no reachable marking puts more than
one token into any place. As explained above, all the nets we are interested in
will be safe. Thus, we shall henceforth treat markings as subsets of~$P$.

An infinite sequence $t_1t_2\ldots$ is called a \emph{run} if every prefix of
it is one. We say that a run $\sigma$ is \emph{fair} iff
\begin{itemize}
\item either $\sigma$ is finite, and in the marking reached by $\sigma$,
  no transition is enabled;
\item or $\sigma=t_1t_2\ldots$ is infinite, where $M_1,M_2,\ldots$ are the
  markings generated by firing $\sigma$, and there exists no pair $t\in T$
  and $i\ge1$ such that $t$ is enabled in all $M_k$, \ $k\ge i$ and
  $t\ne t_k$ for all $k>i$.
\end{itemize}
In other words, a fair run cannot delay firing an enabled transition 
forever.

\subsection{Occurrence nets}

Occurrence nets are a specific type of acyclic Petri net.
Keeping with tradition, we shall call the places of an occurrence net
\emph{conditions} and its transitions \emph{events}.
Fix a safe Petri net $O=(C,E,F,C_0)$ for the rest of this subsection.
We let $<$ denote the transitive closure of $\flow$ and $\le$
the reflexive closure of $<$; further, if $e\in E$ is an event,
let $\cone{e}:=\{\,e'\in E\mid e' \le e\,\}$ be the \emph{cone} of~$e$, and 
$\precone{e}:=\cone{e}\setminus\{e\}$ the \emph{pre-cone} of~$e$. 

Two nodes $x,x'$ are \emph{in conflict}, 
written $x \conflict x'$ if there exist
$e,e'\in E$ such that \emph{(i)} $e\ne e'$, \emph{(ii)} 
$\preset{e}\cap\preset{e'}\ne\emptyset$,  and \emph{(iii)} 
$e\le x$ and $e'\le x'$.

$O$ is called an \emph{occurrence net} if it satisfies the following
properties:
\begin{enumerate}
\item no self-conflict: $\forall x \in C\cup E\colon \neg (x\conflict x)$;
\item $<$ is acyclic, i.e.\ $\le$ is a partial order;
\item finite cones: all events $e$ satisfy $|\cone{e}| < \infty$; 
\item no backward branching: all conditions $c$ satisfy $|\preset{c}|\le1$;
\item $C_0\subseteq C$ is the set of $\le$-minimal nodes.
\end{enumerate}

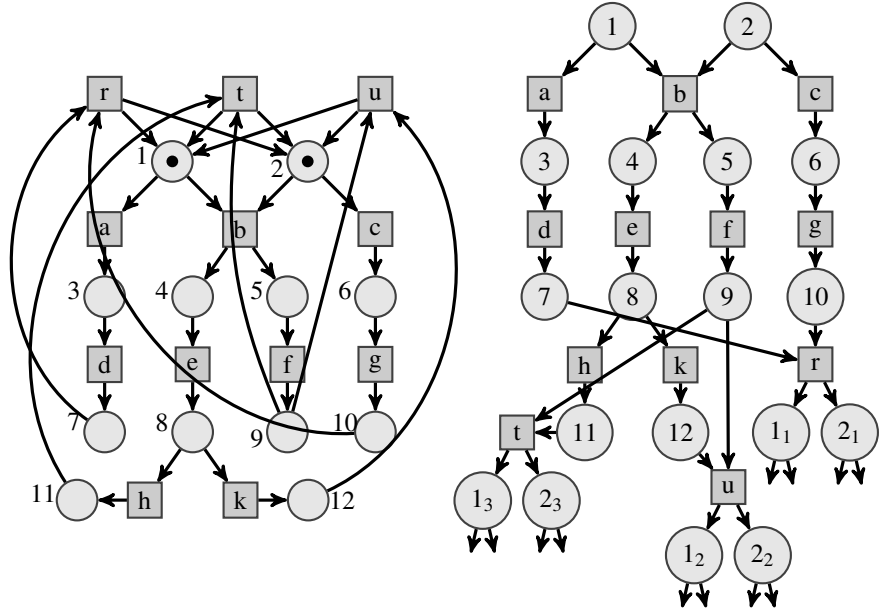
\begin{figure}[htbp]
  \centering
  \begin{tikzpicture}[node distance=1.0cm,>=stealth',bend angle=55,auto,transform shape,yshift=-3cm,scale=0.9]

    \tikzstyle{place}=[circle,thick,draw=black!75,fill=gray!20,minimum size=6mm]
    \tikzstyle{red place}=[place,draw=red!75,fill=red!20]
    \tikzstyle{transition}=[rectangle,thick,draw=black!75,
    fill=black!20,minimum size=5mm]

    \tikzstyle{every label}=[red]
\begin{scope}[yshift=-1cm]
      \node [transition] (rr)  {r};
      \node [] (aa2) [right of=rr] {};
      \node [transition] (tt) [right of=aa2] {t};
      \node [] (aa4) [right of=tt] {};
      \node [transition] (uu) [right of= aa4] {u};

      \node [] (d1) [below of=rr] {};
      \node [place] (f1) [right of=d1,tokens=1] {};
      \node [] [xshift=0.55cm,yshift=0.1cm, left of=f1] {1};
      \node [] (d2) [right of=f1] {};
      \node [place] (f2) [right of=d2,tokens=1] {};
      \node [] [xshift=0.55cm,yshift=-0.1cm, left of=f2] {2};
      \node [] (d3) [right of=f2] {};

      \node [transition] (e1) [below of=d1] {a};
      \node [] (d4) [below of=f1]{};
      \node [transition] (e2) [below of=d2] {b};
      \node [] (d5) [below of= f2 ] {};
      \node [transition] (e3) [below of=d3] {c};

      \node [place] (l1) [below of=e1] {};
      \node [] [xshift=0.55cm,yshift=0.1cm, left of=l1] {3};
      \node [place] (l2) [below of=d4, xshift=3mm] {};
      \node [] [xshift=0.55cm,yshift=0.1cm, left of=l2] {4};
      \node [] (d6) [below of=f2] {};
      \node [place] (l3) [below of=d5, xshift=-3mm] {};
      \node [] [xshift=0.55cm,yshift=0.1cm, left of=l3] {5};
      \node [place] (l4) [below of=e3] {};
      \node [] [xshift=0.55cm,yshift=0.1cm, left of=l4] {6};

      \node [transition] (k1) [below of=l1] {d};
      \node [transition] (k2) [below of=l2] {e};
      \node [] (d7) [below of=d6] {};
      \node [transition] (k3) [below of=l3] {f};
      \node [transition] (k4) [below of=l4] {g};

      \node [place] (m1) [below of=k1] {};
      \node [] [xshift=0.55cm,yshift=0.15cm, left of=m1] {7};
      \node [place] (m2) [below of=k2] {};
      \node [] [xshift=0.55cm,yshift=0.15cm, left of=m2] {8};
      \node [] (d8) [below of=d7] {};
      \node [place] (m3) [below of=k3] {};
      \node [] [xshift=0.55cm,yshift=-0.1cm, left of=m3] {9};
      \node [place] (m4) [below of=k4] {};
      \node [] [xshift=0.55cm,yshift=0.15cm, left of=m4] {10};

      \node [transition] (j1) [below left of=m2,yshift=-3mm] {h};
      \node [transition] (j2) [below right of=m2,yshift=-3mm] {k};

      \node [place] (p1) [left of=j1] {};
      \node [] [xshift=0.5cm,yshift=0.1cm, left of=p1] {11};
      \node [place] (p2) [right of=j2] {};
      \node [] [xshift=-0.5cm,yshift=0cm, right of=p2] {12};

      \draw[->,very thick] (f1) -- (e1);
      \draw[->,very thick] (f1) -- (e2);
      \draw[->,very thick] (f2) -- (e2);
      \draw[->,very thick] (f2) -- (e3);

      \draw[->,very thick] (e1) -- (l1);
      \draw[->,very thick] (e2) -- (l2);
      \draw[->,very thick] (e2) -- (l3);
      \draw[->,very thick] (e3) -- (l4);

      \draw[->,very thick] (l1) -- (k1);
      \draw[->,very thick] (l2) -- (k2);
      \draw[->,very thick] (l3) -- (k3);
      \draw[->,very thick] (l4) -- (k4);

      \draw[->,very thick] (k1) -- (m1);
      \draw[->,very thick] (k2) -- (m2);
      \draw[->,very thick] (k3) -- (m3);
      \draw[->,very thick] (k4) -- (m4);

      \draw[->,very thick] (m2) -- (j1);
      \draw[->,very thick] (m2) -- (j2);

      \draw[->,very thick] (j1) -- (p1);
      \draw[->,very thick] (j2) -- (p2);
      
      \draw[->,very thick] (rr) -- (f1);
      \draw[->,very thick] (rr) -- (f2);
      \draw[->,very thick] (tt) -- (f1);
      \draw[->,very thick] (tt) -- (f2);
      \draw[->,very thick] (uu) -- (f1);
      \draw[->,very thick] (uu) -- (f2);

      \path (m4) 	edge[->,very thick,bend left]  (rr);
      \path (m3) 	edge[->,very thick,bend angle=15,bend left]  (tt);
      \path (p1) 	edge[->,very thick,bend left]  (tt);
      \path (p2) 	edge[->,very thick, bend right]  (uu);
      \path (m3) 	edge[->,very thick]  (uu);
      \path (m1) 	edge[->,very thick,bend left]  (rr);
  \end{scope}
     \begin{scope}[xshift=6.5cm] 
      
      \node [] (d1) {};
      \node [place] (f1) [right of=d1] {1};
      \node [] (d2) [right of=f1] {};
      \node [place] (f2) [right of=d2] {2};
      \node [] (d3) [right of=f2] {};

      \node [transition] (e1) [below of=d1] {a};
      \node [] (d4) [below of=f1]{};
      \node [transition] (e2) [below of=d2] {b};
      \node [] (d5) [below of= f2 ] {};
      \node [transition] (e3) [below of=d3] {c};

      \node [place] (l1) [below of=e1] {3};
      \node [place] (l2) [below of=d4, xshift=3mm] {4};
      \node [] (d6) [below of=f2] {};
      \node [place] (l3) [below of=d5, xshift=-3mm] {5};
      \node [place] (l4) [below of=e3] {6};

      \node [transition] (k1) [below of=l1] {d};
      \node [transition] (k2) [below of=l2] {e};
      \node [] (d7) [below of=d6] {};
      \node [transition] (k3) [below of=l3] {f};
      \node [transition] (k4) [below of=l4] {g};

      \node [place] (m1) [below of=k1] {7};
      \node [place] (m2) [below of=k2] {8};
      \node [] (d8) [below of=d7] {};
      \node [place] (m3) [below of=k3] {9};
      \node [place] (m4) [below of=k4] {10};

      \node [transition] (j1) [below left of=m2,yshift=-3mm] {h};
      \node [transition] (j2) [below right of=m2,yshift=-3mm] {k};

      \node [place] (p1) [below of=j1] {11};
      \node [place] (p2) [below of=j2] {12};
      
      \node [transition] (oo3) [below right of=p2,yshift=-1mm] {u};
      \node [transition] (oo1) [below of=m4] {r};
      \node [transition] (oo2) [left of=p1] {t};

      \node [place] (aj1) [below left of=oo1,yshift=-3mm,xshift=2mm] {1$_1$};
      \node [place] (aj2) [below right of=oo1,yshift=-3mm,xshift=-2mm] {2$_1$};

      \node [place] (bj1) [below left of=oo3,yshift=-3mm,xshift=2mm] {1$_2$};
      \node [place] (bj2) [below right of=oo3,yshift=-3mm,xshift=-2mm] {2$_2$};

      \node [place] (cj1) [below left of=oo2,yshift=-3mm,xshift=2mm] {1$_3$};
      \node [place] (cj2) [below right of=oo2,yshift=-3mm,xshift=-2mm] {2$_3$};

      \node [] (dum1) [below left of=aj1,yshift=-2mm,xshift=5mm] {};
      \node [] (dum2) [below right of=aj1,yshift=-2mm,xshift=-5mm] {};
      \path (aj1) 	edge[->,very thick]  (dum1);
      \path (aj1) 	edge[->,very thick]  (dum2);
      
      \node [] (duma1) [below left of=aj2,yshift=-2mm,xshift=5mm] {};
      \node [] (duma2) [below right of=aj2,yshift=-2mm,xshift=-5mm] {};
      \path (aj2) 	edge[->,very thick]  (duma1);
      \path (aj2) 	edge[->,very thick]  (duma2);
      
      \node [] (dumb1) [below left of=bj1,yshift=-2mm,xshift=5mm] {};
      \node [] (dumb2) [below right of=bj1,yshift=-2mm,xshift=-5mm] {};
      \path (bj1) 	edge[->,very thick]  (dumb1);
      \path (bj1) 	edge[->,very thick]  (dumb2);
      
      \node [] (dumc1) [below left of=bj2,yshift=-2mm,xshift=5mm] {};
      \node [] (dumc2) [below right of=bj2,yshift=-2mm,xshift=-5mm] {};
      \path (bj2) 	edge[->,very thick]  (dumc1);
      \path (bj2) 	edge[->,very thick]  (dumc2);
      
      \node [] (dumd1) [below left of=cj1,yshift=-2mm,xshift=5mm] {};
      \node [] (dumd2) [below right of=cj1,yshift=-2mm,xshift=-5mm] {};
      \path (cj1) 	edge[->,very thick]  (dumd1);
      \path (cj1) 	edge[->,very thick]  (dumd2);
      
      \node [] (dume1) [below left of=cj2,yshift=-2mm,xshift=5mm] {};
      \node [] (dume2) [below right of=cj2,yshift=-2mm,xshift=-5mm] {};
      \path (cj2) 	edge[->,very thick]  (dume1);
      \path (cj2) 	edge[->,very thick]  (dume2);

      \path (m3) 	edge[->,very thick]  (oo2);
      \path (p1) 	edge[->,very thick]  (oo2);
      
      \path (m4) 	edge[->,very thick]  (oo1);
      \path (m1) 	edge[->,very thick]  (oo1);
      
      \path (m3) 	edge[->,very thick]  (oo3);
      \path (p2) 	edge[->,very thick]  (oo3);

      \draw[->,very thick] (oo1) -- (aj1);
      \draw[->,very thick] (oo1) -- (aj2);
      \draw[->,very thick] (oo3) -- (bj1);
      \draw[->,very thick] (oo3) -- (bj2);
      \draw[->,very thick] (oo2) -- (cj1);
      \draw[->,very thick] (oo2) -- (cj2);
      
      \draw[->,very thick] (f1) -- (e1);
      \draw[->,very thick] (f1) -- (e2);
      \draw[->,very thick] (f2) -- (e2);
      \draw[->,very thick] (f2) -- (e3);

      \draw[->,very thick] (e1) -- (l1);
      \draw[->,very thick] (e2) -- (l2);
      \draw[->,very thick] (e2) -- (l3);
      \draw[->,very thick] (e3) -- (l4);

      \draw[->,very thick] (l1) -- (k1);
      \draw[->,very thick] (l2) -- (k2);
      \draw[->,very thick] (l3) -- (k3);
      \draw[->,very thick] (l4) -- (k4);

      \draw[->,very thick] (k1) -- (m1);
      \draw[->,very thick] (k2) -- (m2);
      \draw[->,very thick] (k3) -- (m3);
      \draw[->,very thick] (k4) -- (m4);

      \draw[->,very thick] (m2) -- (j1);
      \draw[->,very thick] (m2) -- (j2);

      \draw[->,very thick] (j1) -- (p1);
      \draw[->,very thick] (j2) -- (p2);
     \end{scope} 

  \end{tikzpicture}
     \caption{{A Petri net (left) and a prefix of its unfolding (right)}}
    \label{fig:netandon}
    \label{fig:U_complete}
\end{figure}

\begin{example}
The right hand side of Figure \ref{fig:netandon} shows an occurrence net. The events $a$ and $c$ are both in conflict
with $b$, yet not with one another; in fact, they are \emph{concurrent} (neither ordered nor in conflict). 
\end{example}

Let $O=(C,E,F,C_0)$ be an occurrence net.
We call $O'=(C',E',F',C_0)$ a \emph{prefix} of~$O$ if
\begin{itemize}
\item $C'\subseteq C$, \ $E'\subseteq E$, \ $F'=F\cap (C'\cup E')^2$,
  and moreover $C'\supseteq C_0\cup\postset{(E')}$;
\item $C'$ and $E'$ are downward-closed, i.e.\ for any $x\in C'\cup E'$ and
  $y<x$ we have $y\in C'\cup E'$.
\end{itemize}
A prefix is called finite if $C'$ and $E'$ are finite sets.
Notice that each prefix is uniquely determined by its set of events.
We denote by $O[E']$ the unique prefix of~$O$ whose set of events is~$E'$.

Let $\C\subseteq E$ be a downward-closed and conflict-free set of events,
that is, $e\in\C$ and $e'<e$ imply $e'\in\C$, and $e,e'\in\C$ implies
$\neg(e\conflict e')$. Then we call $\C$ a \emph{configuration} of~$O$.
Given a configuration~$\C$, we define $\Cut(\C)$ to be the set of 
$\le$-maximal conditions of $O[\C]$. Moreover we define the
\emph{postfix} $\postfix{O}{\C}$ to be the occurrence net
$(C'',E'',F'',C_0'')$, where
$C''=C\setminus\preset{\C}$, \ $E''=E\setminus\C$, $F''=F\cap(C''\cup E'')^2$, and $C_0''=\Cut(\C)$.

If $\C$ is a finite configuration and
$e\in E\setminus\C$ an event such that $\preset{e}\subseteq\Cut(\C)$.
In this case, $\C':=\C\cup\{e\}$ is a configuration, and we
write $\C\omove{e}$ or $\C\omove{e}\C'$. By extension, 
for a finite configuration $\C$ and a set $A=\{e_1,\ldots,e_n\}$
of events, we write $\C\omove{A}\C'$ iff there exist $\C_0,\ldots,\C_n$ such that
$\C_0=\C$, \ $\C_n=\C'$, and for all $i=1,\ldots,n$,
  $\C_{i-1}\omove{e_i}\C_i$. We write $\C\emogl\C'$ if there exists a set~$A$
such that $\C\omove{A}\C'$.

The following facts are well-known, see e.g.~\cite{BPEng,esparza-1}:
\begin{itemize}
\item A downward-closed set $\C\subseteq E$ is a configuration iff
  the elements of $\C$ can be arranged to form a run $\sigma$ of $O$.
  We have that $\sigma$ is fair iff $\C$ is maximal.
  Moreover, if $\C$ is finite, then $\sigma$ leads from $C_0$ to $\Cut(\C)$.
\item For every event $e$, $\cone{e}$ and $\precone{e}$ are configurations.
\item Let $c,c'\in C$ be a pair of conditions. Then exactly one of the following
  three statements holds:
\begin{itemize}
\item $c$ and $c'$ are \emph{causally related}, i.e.\ $c<c'$ or $c'<c$;
\item $c$ and $c'$ are in \emph{conflict}, i.e.\ $c\conflict c'$;
\item $c$ and $c'$ are called \emph{concurrent}, written $c\concurrent c'$,
  i.e.\ there exists a configuration $\C$ such that $\{c,c'\}\subseteq\Cut(\C)$.
\end{itemize}
A set of pairwise concurrent places is called a co-set.
\end{itemize}

\subsection{Unfoldings}
Let ${N=(P,T,F,M_0)}$ be a safe Petri net. Intuitively, an unfolding of $N$ is
an acyclic version of $N$ where loops of $N$ are
``unrolled''; an unfolding is usually infinite even if $N$ is finite.

Formally, $U=(C,E,G,C_0)$ is called an \emph{unfolding} of $N$ if $U$
is an occurrence
net equipped with a mapping $f\colon(C\cup E)\to(P\cup T)$, which we extend
to sets and sequences in the usual way. We shall write $f\bijects{A}{B}$
if the restriction of $f$ to $A$ yields a bijection between $A$ and $B$.
Then $U$ is the unfolding of $N$ if the following properties hold:
\begin{itemize}
\item $f(C)\subseteq P$, \ $f(E)\subseteq T$, and $f\bijects{C_0}{M_0}$;
\item for every co-set 
  $D\subseteq C$ and transition~$t\in T$ such that
  $f\bijects{D}{\preset{t}}$, there is exactly one event $e\in E$ with
  $f(e)=t$ and $\preset{e}=D$;
\item if $f(e)=t$ for some event~$e$, then $f\bijects{\preset{e}}{\preset{t}}$ and $f\bijects{\postset{e}}{\postset{t}}$.
\end{itemize}

With every configuration $\C$ of~$U$ we associate the marking
$\Mark(\C):=\{\,f(c)\mid c\in\Cut(\C)\,\}$.

\begin{example}
Figure \ref{fig:netandon} shows a net $N$ on the left and prefix
of its unfolding on the right; the function~$f$ is reflected in 
the inscriptions. 
It is well-known~\cite{BPEng,esparza-1} that $M$ is a reachable marking in $N$
iff there exists a configuration $\C$ of $U$ such that
$\Mark(\C)=M$.
Moreover, if $\sigma$ is a run corresponding to $\C$, then $f(\sigma)$
leads from $M_0$ to $M$ in $N$. It is in this sense that $U$ mimics
the behaviour of $N$.
\end{example}

A prefix $U'$ of $U$ is called \emph{complete} if it ``contains'' every marking
of $N$, i.e. for every reachable marking $M\in\reach(N)$ there exists
a configuration~$\C$ of $U'$ such that $\Mark(\C)=M$.
It is well-known that for any configuration~$\C$,
the postfix $\postfix{U}{\C}$ is isomorphic to the unfolding of the net
$(P,T,F,\Mark(\C))$.

\subsection{The ``reveals'' relation}

\begin{figure}[htbp]
  \centering
  \begin{tikzpicture}[node distance=0.9cm,>=stealth',bend angle=45,auto]

    \tikzstyle{place}=[circle,thick,draw=black!75,fill=gray!20,minimum size=6mm]
    \tikzstyle{red place}=[place,draw=red!75,fill=red!20]
    \tikzstyle{transition}=[rectangle,thick,draw=black!75,
    fill=black!20,minimum size=5mm]

    \tikzstyle{every label}=[red]

    \begin{scope}
      \node [] (d1) {};
      \node [place] (f1) [right of=d1] {1};
      \node [] (d2) [right of=f1] {};
      \node [place] (f2) [right of=d2] {2};
      \node [] (d3) [right of=f2] {};

      \node [transition] (e1) [below of=d1] {a};
      \node [] (d4) [below of=f1]{};
      \node [transition] (e2) [below of=d2] {b};
      \node [] (d5) [below of= f2 ] {};
      \node [transition] (e3) [below of=d3] {c};

      \node [place] (l1) [below of=e1] {3};
      \node [place] (l2) [below of=d4, xshift=3mm] {4};
      \node [] (d6) [below of=f2] {};
      \node [place] (l3) [below of=d5, xshift=-3mm] {5};
      \node [place] (l4) [below of=e3] {6};

      \node [transition] (k1) [below of=l1] {d};
      \node [transition] (k2) [below of=l2] {e};
      \node [] (d7) [below of=d6] {};
      \node [transition] (k3) [below of=l3] {f};
      \node [transition] (k4) [below of=l4] {g};

      \node [place] (m1) [below of=k1] {7};
      \node [place] (m2) [below of=k2] {8};
      \node [] (d8) [below of=d7] {};
      \node [place] (m3) [below of=k3] {9};
      \node [place] (m4) [below of=k4] {10};

      \node [transition] (j1) [below left of=m2,yshift=-3mm] {h};
      \node [transition] (j2) [below right of=m2,yshift=-3mm] {k};

      \node [place] (p1) [below of=j1] {11};
      \node [place] (p2) [below of=j2] {12};

      \draw[->,very thick] (f1) -- (e1);
      \draw[->,very thick] (f1) -- (e2);
      \draw[->,very thick] (f2) -- (e2);
      \draw[->,very thick] (f2) -- (e3);

      \draw[->,very thick] (e1) -- (l1);
      \draw[->,very thick] (e2) -- (l2);
      \draw[->,very thick] (e2) -- (l3);
      \draw[->,very thick] (e3) -- (l4);

      \draw[->,very thick] (l1) -- (k1);
      \draw[->,very thick] (l2) -- (k2);
      \draw[->,very thick] (l3) -- (k3);
      \draw[->,very thick] (l4) -- (k4);

      \draw[->,very thick] (k1) -- (m1);
      \draw[->,very thick] (k2) -- (m2);
      \draw[->,very thick] (k3) -- (m3);
      \draw[->,very thick] (k4) -- (m4);

      \draw[->,very thick] (m2) -- (j1);
      \draw[->,very thick] (m2) -- (j2);

      \draw[->,very thick] (j1) -- (p1);
      \draw[->,very thick] (j2) -- (p2);
    \end{scope}

  \end{tikzpicture}
  \caption{Example of an occurrence net}\label{fig:example}
\label{fig:facets}
\end{figure}
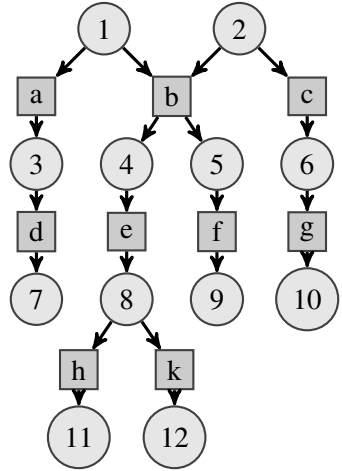

To illustrate ``reveals'' we shall study the occurrence net
in Figure~\ref{fig:facets}. We are interested in finding relations
between events of the form
'if $x$ occurs, then $y$ has already occurred, or will occur eventually',
in the sense that any fair run that contains $x$ also contains $y$.
In other words, this means that \emph{$y$ is inevitable given $x$}.

In the context of Figure~\ref{fig:example}, it is obvious that, 
for any fair run $\sigma$, 
$$k\in\sigma \implies e\in\sigma \implies b\in\sigma,$$
where we use $k\in\sigma$ etc informally to mean that $k$ occurs somewhere
in $\sigma$.  In fact, the statement above simply reflects the causal
relationship; if $k$ happens, then surely its cause $e$ must have happened
before.

But one also obtains the following facts in Figure \ref{fig:example},
again for fair runs $\sigma$:
$$a\in\sigma\iff \neg(b\in\sigma)\iff c\in\sigma
\qquad\hbox{and}\qquad
c\in\sigma\iff g\in\sigma.$$
In fact, $a,c$ are a pair of independent transitions which can happen
concurrently, where as $c$ is a causal predecessor of $g$ and yet allows to
determine that $g$ will eventually happen.
The reader is invited to check that these relations
follow from the fairness of runs.
We thus define our desired relation as follows:

\begin{definition}
Let $O$ be an
occurrence net and $e,e'$ be two of its events. We say that
$e$ reveals $e'$, written $e\reveals e'$, iff for all fair runs $\sigma$ of $O$
$e\in\sigma$ implies $e'\in\sigma$.
The \emph{revealed range} of event $e$ is $\cov{e}:=\{\,e'\mid e\reveals e'\,\}$.
\end{definition}

Notice that the definition immediately implies that $\reveals$ is reflexive and
transitive. Moreover, there is a reveals relationship along causal successors,
i.e.\ if $a<b$, then $b\reveals a$.
The relation $\reveals$ is not symmetric in general: in fact, in
Figure~\ref{fig:example} we have $h\reveals e$ but $\neg(e\reveals h)$.
On the other hand, $\reveals$
is not a partial order: consider $e\reveals f$ and $f\reveals e$
in Figure~\ref{fig:example}.

These examples show that the inheritance of conflict along causality
relations is not sufficient to derive the statements above.
One might therefore suspect that, to obtain the above facts
one would have to explore the entire set of configurations.
However, the following is known:
\begin{lemma}[\cite{CDC07,Haar-tac10}]
\label{lemma:witness}
For an event $e$,
its \emph{conflict set} is defined as $\flict{e}:=\{\,e'\mid e\conflict e'\,\}$.
We have that $e\reveals e'$ iff $\flict{e}\supseteq\flict{e'}$.
\end{lemma}
 
Thus, in principle all it takes to see if $e\reveals e'$ holds is to
check whether no \emph{witness} against it exists for $(e,e')$;
we call $g$ a \emph{witness} for the tuple $(e,e')$ if 
$\neg(e\conflict g)$ and $e'\conflict g$.
However, notice that this does not provide us with an effective procedure
because the conflict sets can be infinite in general (see \cite{Haar-tac10}). 
In Section~\ref{sec:reveal} we shall  show
that $e\reveals e'$ can effectively be decided.

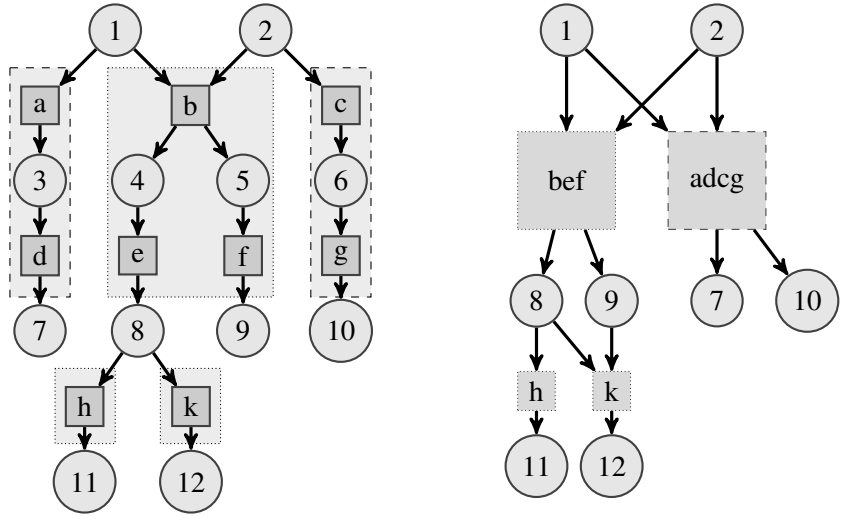
\begin{figure}[htbp]
  \centering
  \begin{tikzpicture}[node distance=1.0cm,>=stealth',bend angle=45,auto]

    \tikzstyle{place}=[circle,thick,draw=black!75,fill=gray!20,minimum size=6mm]
    \tikzstyle{red place}=[place,draw=red!75,fill=red!20]
    \tikzstyle{transition}=[rectangle,thick,draw=black!75,
    fill=black!20,minimum size=5mm]

    \tikzstyle{every label}=[red]

    \begin{scope}
      \draw[densely dotted, fill=gray!30,fill opacity=0.5] (0.9,-0.5) rectangle (3.1, -3.55);
      \draw[dashed, fill=gray!30,fill opacity=0.5] (-0.4,-0.5) rectangle (0.4, -3.55);
      \draw[dashed, fill=gray!30,fill opacity=0.5] (3.6,-0.5) rectangle (4.4, -3.55);
      \draw[densely dotted, fill=gray!30,fill opacity=0.5] (0.2,-4.5) rectangle (1.0, -5.5);
      \draw[densely dotted, fill=gray!30,fill opacity=0.5] (1.6,-4.5) rectangle (2.4, -5.5);
      \node [] (d1) {};
      \node [place] (f1) [right of=d1] {1};
      \node [] (d2) [right of=f1] {};
      \node [place] (f2) [right of=d2] {2};
      \node [] (d3) [right of=f2] {};

      \node [transition] (e1) [below of=d1] {a};
      \node [] (d4) [below of=f1]{};
      \node [transition] (e2) [below of=d2] {b};
      \node [] (d5) [below of= f2 ] {};
      \node [transition] (e3) [below of=d3] {c};

      \node [place] (l1) [below of=e1] {3};
      \node [place] (l2) [below of=d4, xshift=3mm] {4};
      \node [] (d6) [below of=f2] {};
      \node [place] (l3) [below of=d5, xshift=-3mm] {5};
      \node [place] (l4) [below of=e3] {6};

      \node [transition] (k1) [below of=l1] {d};
      \node [transition] (k2) [below of=l2] {e};
      \node [] (d7) [below of=d6] {};
      \node [transition] (k3) [below of=l3] {f};
      \node [transition] (k4) [below of=l4] {g};

      \node [place] (m1) [below of=k1] {7};
      \node [place] (m2) [below of=k2] {8};
      \node [] (d8) [below of=d7] {};
      \node [place] (m3) [below of=k3] {9};
      \node [place] (m4) [below of=k4] {10};

      \node [transition] (j1) [below left of=m2,yshift=-3mm] {h};
      \node [transition] (j2) [below right of=m2,yshift=-3mm] {k};

      \node [place] (p1) [below of=j1] {11};
      \node [place] (p2) [below of=j2] {12};

      \draw[->,very thick] (f1) -- (e1);
      \draw[->,very thick] (f1) -- (e2);
      \draw[->,very thick] (f2) -- (e2);
      \draw[->,very thick] (f2) -- (e3);

      \draw[->,very thick] (e1) -- (l1);
      \draw[->,very thick] (e2) -- (l2);
      \draw[->,very thick] (e2) -- (l3);
      \draw[->,very thick] (e3) -- (l4);

      \draw[->,very thick] (l1) -- (k1);
      \draw[->,very thick] (l2) -- (k2);
      \draw[->,very thick] (l3) -- (k3);
      \draw[->,very thick] (l4) -- (k4);

      \draw[->,very thick] (k1) -- (m1);
      \draw[->,very thick] (k2) -- (m2);
      \draw[->,very thick] (k3) -- (m3);
      \draw[->,very thick] (k4) -- (m4);

      \draw[->,very thick] (m2) -- (j1);
      \draw[->,very thick] (m2) -- (j2);

      \draw[->,very thick] (j1) -- (p1);
      \draw[->,very thick] (j2) -- (p2);
    \end{scope}
    \begin{scope}[xshift=6cm]
      \node [] (d1) {};
      \node [place] (f1) [right of=d1] {1};
      \node [] (d2) [right of=f1] {};
      \node [place] (f2) [right of=d2] {2};
      \node [] (d3) [right of=f2] {};
      \node[rectangle,draw=black!75, densely dotted, fill=gray!30,minimum size=1.3cm,below of=f1,yshift=-1cm](g1){bef};
      \node[rectangle,draw=black!75, dashed, fill=gray!30,minimum size=1.3cm,below of=f2,yshift=-1cm](g2){adcg};
      \node [place] (m1) [below of=g1, yshift=-0.6cm, xshift=-0.4cm] {8};
      \node [place] (m2) [below of=g1, yshift=-0.6cm, xshift=0.6cm] {9};
      \node [] (d8) [below of=d7] {};
      \node [place] (m3) [below of=g2, yshift=-0.6cm, xshift=0cm] {7};
      \node [place] (m4) [below of=g2, yshift=-0.6cm, xshift=1.2cm] {10};
      \node[rectangle,draw=black!75, densely dotted, fill=gray!30,minimum size=0.5cm,below of=m1,yshift=-0.2cm](l1){h};
      \node[rectangle,draw=black!75, densely dotted, fill=gray!30,minimum size=0.5cm,below of=m2,yshift=-0.2cm](l2){k};
      \node [place] (p1) [below of=l1] {11};
      \node [place] (p2) [below of=l2] {12};
      \draw[->,very thick] (f1) -- (g1);
      \draw[->,very thick] (f1) -- (g2);
      \draw[->,very thick] (f2) -- (g1);
      \draw[->,very thick] (f2) -- (g2);
      \draw[->,very thick] (g1) -- (m1);
      \draw[->,very thick] (g1) -- (m2);
      \draw[->,very thick] (g2) -- (m3);
      \draw[->,very thick] (g2) -- (m4);
      \draw[->,very thick] (m1) -- (l2);
      \draw[->,very thick] (m1) -- (l1);
      \draw[->,very thick] (m2) -- (l2);
      \draw[->,very thick] (l1) -- (p1);
      \draw[->,very thick] (l2) -- (p2);
    \end{scope}

  \end{tikzpicture}
  \caption{Left: a prefix of the example from Fig.  \ref{fig:facets}
  with facets highlighted; right: the occurrence net obtained from the 
  left hand one through facet abstraction}
  \label{fig:Facets4}	\label{fig:Facets1}
\end{figure}

\paragraph*{Facets.}
\label{sub:facets}
Let us just note in passing that the strongly connected components of
$\reveals$, called  \emph{facets} in \cite{Haar-tac10}, form 
 equivalence class of occurrence in the sense that any run $\omega$ that contains \emph{any} event of a facet must contain \emph{all} of its 
events. 
In Figure~\ref{fig:Facets4}, the decomposition of the occurrence net from
Figure~\ref{fig:facets} into its facets is shown.
The facets are
$\{a,d,c,g\},\ \{b,e,f\},\  \{h\},\ \{k\}$; 
the right hand side shows the occurrence net obtained by abstracting every facet
into a single event. In general, quotienting an occurrence net into its facets
and their boundary conditions
yields an occurrence net whose set of maximal runs is in bijection with that of the initial occurrence net;
this procedure (for details see \cite{Haar-tac10}) can reduce the model size for analyses of any properties regarding \emph{maximal} behaviours. In \cite{BalChaHaa11}, we focus on \emph{reduced} nets,
i.e. where the contraction of facets has been carried out, and 
every event is a facet; in this framework, behavioural properties 
can be specified in a dedicated logic ERL, for which the synthesis
problem is solved in \cite{BalChaHaa11}; the occurrence nets 
obtained in a canonical way from a logical formula belong to a distinguished subclass of reduced occurrence nets, the \emph{tight} nets. 
For more traditional applications, the facet decomposition can in general yield fast sufficient
criteria for verifying properties. Consider observability-related properties  Petri nets (see \cite{CDC07,CDC09} for a detailed discussion on diagnosability): if $\lambda:\trans\to A$ is a \emph{partial} labelling in some alphabet $A$, how can one quickly decide whether some \emph{unobservable}
transition $t$ - i.e. on which $\lambda$ is undefined - has occured? By pre-computing the reveals-relation and thus the facets on a sufficient finite prefix of the unfolding, online reasonings of the following type become available : If $\lambda$ is such that every facet
in which some instance of  $t$ occurs contains an occurrence of a distinctive label $a$ that $t$ free facets do not produce, then detection of $a$ allows to infer occurrence of $t$ with certainty. 
Given that the facet decomposition and contraction 
can be computed offline, see below, and reduces the size of unfoldings dramatically, such improvements are 
valuable in monitoring and supervising large distributed networks, in particular in telecommunications
\cite{cantavoid,asynchdiag,distdiag}.

\section{A bound for deciding the \emph{reveals} relation}
\label{sec:reveal}

Let $N=(P,T,F,M_0)$ be a safe Petri net, where $P$ and $T$ are finite,
for the rest of the section, and let $U=(C,E,G,C_0)$ be its unfolding,
where $f$ is the mapping between $U$ and $N$.

In this section, we shall consider
the following problem: Given two events $x$ and $y$, does $x$ reveal~$y$?
As pointed out in Lemma~\ref{lemma:witness}, this requires to decide
whether a witness exists. We shall show that the height of a witness is
bounded, i.e.\ it suffices
to search a finite prefix of $U$ to find a witness. The existence of a
finite bound, albeit a much higher one, was first pointed out
in~\cite{Haar-tac10}, and we start by re-stating that result.

\begin{definition}
\label{def:cutoffs}
Associate to each event $e$ a marking of $\netn$ by taking
$M_e:=\Mark(\cone{e})$.
We shall define a sequence $(L_i)_{i\ge1}$ of sets of events, the so-called
\emph{level-$i$ cutoffs}, and a sequence of prefixes $(U_i)_{i\ge1}$, the
so-called \emph{level-$i$ prefixes}.

We let $e\in L_1$ if $M_e=M_0$ or
there exists an event $e'$ such that $e'<e$ and $M_{e'}=M_e$.
For $i>1$, we let $e\in L_i$ iff there exists an event
$e'\in L_{i-1}$ such that $e'<e$ and $M_{e'}=M_e$.
For $i\ge1$, let $L_i^{\min}$ be the $\le$-minimal events of $L_i$.
We let $U_i:=U[L_i']$, where $L_i':=\bigcup_{e\in L_i^{\min}}\cone{e}$
is the downward-closure of $L_i^{\min}$.
\end{definition}

Intuitively, the prefix $U_1$ contains all reachable markings and unrolls each
loop in the Petri net exactly once; notice that the events $L_1$ are exactly
those events that return the net to a marking that was reached before. The
prefix $U_2$ unrolls each loop once more and so on.
The following result is shown in~\cite{Haar-tac10}:
\bet\label{th:oldresult} \cite{Haar-tac10}
Let $m$ be the
the minimal index such that $U_m$ contains event~$x$, and let $n$
be the corresponding index for $y$. Moreover, let $K_M$ be the number
of reachable markings of the net $\netn$. Then, if $\neg(x\reveals y)$,
there exists a witness in $U_{K_M+\max\{m,n\}-1}$.
\ent

$K_M$ is guaranteed to be finite for safe nets, hence Theorem~\ref{th:oldresult}
establishes the decidability of $\reveals$. However, $K_M$ is difficult to
determine exactly and in general very large, not to mention the size of
$U_{K_M+\max\{m,n\}-1}$. We shall see that this bound can be improved.
Formalizing the discussion after Lemma~\ref{lemma:witness},
we define, for events $x,y,z$, the \emph{witness predicate} $\witness(x,y,z)$:
$$
\witness(x,y,z)\quad:\Longleftrightarrow\quad
\left(z\conflict y\right)\ \land \ \neg\left(z\conflict x\right).
$$
To prepare the main result, let us first define the \emph{height function}
$\heightsym$. Let $O$ be an occurrence net and $e$ one of its events.
Then
$$\height{e}:=1+\max_{e'\in\preset{(\preset{e})}}\height{e'},
\qquad\hbox{where $\max\emptyset:=0$}.$$
We naturally extend the height function to finite prefixes of~$O$:
\begin{eqnarray}
\label{eq:prefhei}
\height{O[E']}:=\max_{e\in E'}\height{e}
\end{eqnarray}

Let $M$ be a reachable marking of $\netn$ and $\netn(M)$ be the net
$(P,T,F,M)$, i.e.\ $\netn$ with $M$ as the initial marking. Moreover,
let $U^M$ be the unfolding of $\netn(M)$ and $U^M_i$ the analogous
prefixes according to Definition~\ref{def:cutoffs}.
Let $K(M):=\height{U^M_1}$, and
\begin{eqnarray}
\label{eqn:K}
K:=\max_{M\in\reach(N)} K(M).
\end{eqnarray}

\begin{lemma}\label{le:U2}
The value of $K$ is bounded above by the height $\height{U_2}$ of 
the level-2 prefix of $N$.
\end{lemma}
\bepr
We first show that $U_1$ is a complete prefix. Indeed, in~\cite{McMillan}
an event $e$ is called a cut-off of $U$ if $M_e=M_0$ or there exists
an event $e'$ such that $M_{e'}=M_{e}$ and $|\cone{e'}|<|\cone{e}|$.
It is shown in~\cite{McMillan} that a prefix that contains all minimal
cutoffs is complete.
Evidently, $e'<e$ implies $|\cone{e'}|<|\cone{e}|$ and is a stronger
condition, therefore our prefix $U_1$ contains all such minimal cutoffs
and is also complete.

Let $M\in\reach(N)$. By completeness of $U_1$,
there exists a configuration $\C$ in $U_1$ such that
$\Mark(\C)=M$. Now, by construction of $U_2$,
the postfix $\postfix{U_2}{\C}$ contains an isomorphic copy of $U_1^M$.
\eepr

\smallskip

\begin{figure}[!ht]
 \centering
\includegraphics[width=0.66\textwidth]{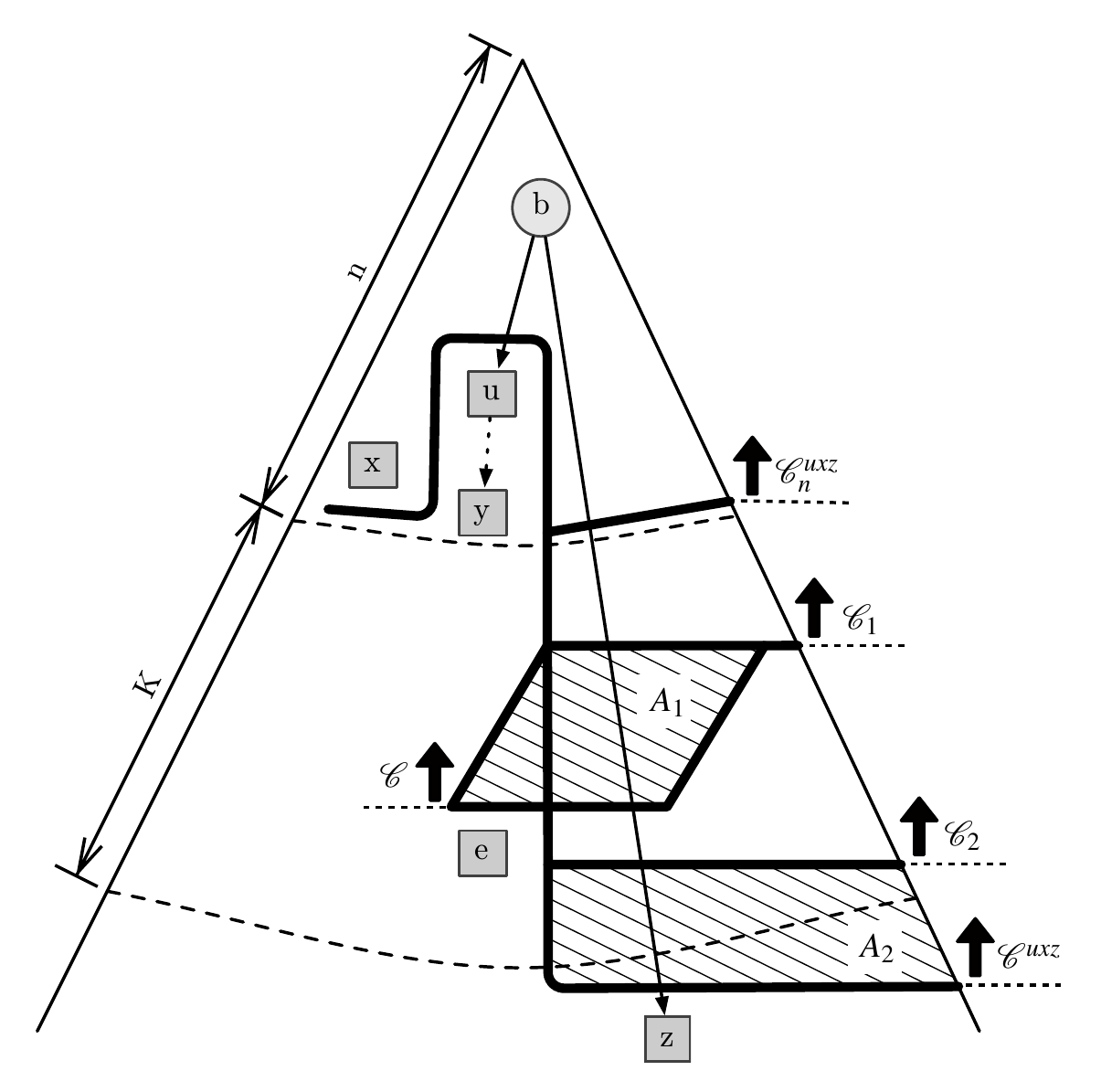}
  \caption{Rough sketch of the proof of Theorem~\ref{th:facetscomputable};
  there exists a condition $b$ in the preset of both $u$ and $z$;
  moreover, $u<y$ and $n=\max(\height{x},\height{y})$. From $\C^{uxz}$ we
  construct the smaller configuration $\C$.}\label{fig:proof}
\end{figure}

We now state the main result of this section:
\begin{theorem}
\label{th:facetscomputable} 
Let $N$ be a safe Petri net, $U$ its unfolding,
and let $K$ as defined in~(\ref{eqn:K}).
For any two events $x,y$ such that $\neg(x\reveals y)$,
there exists an event~$z$ such that
\begin{enumerate}
\item $\witness(x,y,z)$ and
\item $\height{z}\leq n+K$, where $n:=\max(\height{x},\height{y})$.
\end{enumerate}
\end{theorem}

\bepr
The idea of the proof is illustrated in Figure \ref{fig:proof}.
Let $f$ be the mapping between $N$ and $U$.
If $\neg(x\reveals y)$ then some event $z$ satisfying $\witness(x,y,z)$ exists;
it remains to determine the maximal height of $z$.
If $x\conflict y$, we are done immediately, taking $z:=x$.
Otherwise, $\C_{xy}:=\cone{x}\cup\cone{y}$ is a configuration.
Choose $z\in E$ such that $\witness(x,y,z)$
holds, and such that $z'<z$ implies $\neg\witness(x,y,z')$.
By assumption we have $\neg(x\conflict z)$, thus
$\C_{xz}:=\cone{x}\cup\cone{z}$ is also a configuration.
Further, let $u$ be such that $u\conflict z$ and $u\le y$ and such that
$u'<u$ implies $\neg (u'\conflict z)$.
We claim that 
\begin{eqnarray*}
\C^{uxz}&\define&\precone{u}\cup\cone{x}\cup\precone{z}
\end{eqnarray*}
is a configuration:
if this were not the case, then there would be events
$e,e'\in\C^{uxz}$ such that $e\conflict e'$. Since $\C_{xy}$ and $\C_{xz}$
are configurations,
it would follow w.l.o.g.\ that $e\in\precone{u}$ and $e'\in\precone{z}$, so
$e<u$ and $e'<z$. But then $e\conflict z$ and $e'\conflict y$, both of which
contradicts the minimality assumptions on $u$ and $z$. We thus have
\begin{eqnarray}
\label{eqn:Cuxz}
\C^{uxz}\omove{z}&\hbox{and}&
\C^{uxz}\omove{u}\ .
\end{eqnarray}
For $n=\max\{\height{x},\height{y}\}$, let
$\C^{uxz}_n:=\{\,e\in\C^{uxz}\mid\height{\otransn}\le n\,\}$. 
Then $x\in \C^{uxz}_n$, and  $\C^{uxz}_n\omove{u}$.
Suppose that $z$ satisfies $\height{z}>n+K$.
Then the choice of~$K$ implies the existence of two distinct
configurations $\C_1,\C_2$ of $U$ such that

\begin{enumerate}
\item $\C^{uxz}_n\emogl\C_1\emogl\C_2\emogl\C^{uxz}$,
\item $\height{\C_1}<\height{\C_2}$, and
\item $\Mark(\C_1)=\Mark(\C_2)$.
\end{enumerate}

In fact, $\Mark(\C_1)=\Mark(\C_2)$ implies that
$\postfix{U}{\C_1}$ and $\postfix{U}{\C_2}$ are isomorphic, and
there exist sets $A_1$, $A_2$ with $f(A_1)=f(A_2)$ such that
$\C_2\omove{A_2}\C^{uxz}$ and
$\C_1\omove{A_1}\C$ for some $\C$.
Now, $\Mark(\C)=\Mark(\C^{uxz})$, so there exists
an event~$e$ such that
$f(e)=f(z)$, $\height{e}<\height{z}$, and $\C\omove{e}$.
Thus, $\C\cup\{e\}$ is a configuration containing both $x$ and $e$,
so $\neg(x\conflict e)$.

From $u\conflict z$ and (\ref{eqn:Cuxz}) it follows that $u$ and $z$
compete directly for a token, i.e.\ there exists a
condition~$b\in\preset{u}\cap\preset{z}$. Since $f(e)=f(z)$, there must
be $b'\in\preset{e}$ with $f(b')=f(b)$. Now, $b\concurrent b'$ cannot
hold because $N$ is safe. Suppose $b\conflict b'$. But then there must
exist two events $u'\ne e'$ such that $u'<b$ and $e'<b'$ and
$\preset{u'}\cap\preset{e'}\ne\emptyset$. By definition,
$\C$ contains $\precone{u}$ and enables~$e$, so
$b$ and $b'$ must both be contained in the prefix $U[\C]$, so
$u',e'\in\C$, but, being a configuration, $\C$ cannot contain two
conflicting events.
The only possibilities left are $b=b'$, $b<b'$, or $b'<b$, and in all
cases we obtain $e\conflict u$ and therefore $e\conflict y$.

We thus obtain $\witness(x,y,e)$, and the
height of~$e$ is strictly less than that of~$z$.
Either $\height{e}\le n+K$, and we are done; or we replace
$z$ by $e$ and repeat the surgery above, obtain another witness with strictly
lesser height etc, until we end up with a witness that has
the desired height.
\eepr

Theorem~\ref{th:facetscomputable} in connection with
Lemma~\ref{le:U2} implies that for any pair $x,y$ of concurrent events, 
it suffices to inspect $U_2^{M_{xy}}$ to determine whether $x\reveals y$, where 
$M_{xy}:=M(\cone{x}\cup\cone{y})$. Notice that this bound is much
lower than the one given by Theorem~\ref{th:oldresult}; in fact, contrary
to the previous bound it provides hope to actually compute the relation.

The reader will observe that in the proof of Theorem~\ref{th:facetscomputable}
we exploit the fact that a suffix of $\C_n^{uxz}$ with height
$K$ contains two marking-equivalent causally related events. To find two such
events, it actually suffices to search an isomorphic copy of the level-1 prefix
starting at the marking associated with $\C_n^{uxz}$. It is thus
tempting to think that Lemma~\ref{le:U2} unfolds ``one level too much''.
However, for a given candidate $z$ as witness for $x$ and $y$, there may be
many possible events~$u$ for which one would have to search the suffix of
$\C_n^{uxz}$, therefore limiting the candidates in this manner would not
at all be straightforward. The value of Lemma~\ref{le:U2} is in bounding the
set of candidates for $z$ in a simple, effective manner.

\section{Algorithms for computing the \emph{reveals} relation}
\label{sec:alg}

In this section, we exploit the results of Sections \ref{sec:def}
and~\ref{sec:reveal} to exhibit two concrete algorithms for determining
the \emph{reveals} relation. The main contribution is in Section~\ref{sec:fullrel},
where we show how to compute the relation between all events in
a given prefix. In Section~\ref{sec:pair} we discuss the question how
to decide $x\reveals y$ for a single pair $x,y$.

\subsection{Computing \emph{reveals} on a given prefix}
\label{sec:fullrel}

For the rest of this section, let us fix a finite occurrence net $O$,
which should be a finite prefix of some safe Petri net, where $E$ is the
set of events. We are going to compute the relation $\reveals$ between all
pairs in~$E$.

An algorithm for this purpose can be useful if either the underlying
net is free of loops (and hence the unfolding is finite), or if one
wants to compute the relation for all events of height up to
$n$ (in which case the prefix should contain the events of height $n+K$).

Our algorithm consists of three passes over the occurrence net that compute, in
turn, the causality relation $<$, the conflict relation $\conflict$,
and finally the reveals relation $\reveals$. We assume that events in
$E$ are available in topologically sorted order, i.e.\ an order $\prec$
where $e<e'$ implies $e\prec e'$. Such an order
can be easily established while scanning $O$:
e.g., one first identifies the minimal conditions (those having no incoming
arcs) and then traverses the unfolding with a standard worklist algorithm.

For the three passes that compute $<$, $\conflict$, and $\reveals$, we
exploit certain causal inheritance properties. It turns out that most
operations can be implemented with simple bitset operations.
\begin{enumerate}
\item In the first pass, we compute for each event $e$ a set of events
  $\post{e}:=\{\,e'\mid e\le e'\,\}$ containing its successors (and $e$
  itself). Initially, that set is empty for all $e$; we then traverse
  $E$ in \emph{inverse} topological order, exploiting the fact
  that the causal relationship is obviously transitive: $e\le e'$ iff
  $e=e'$ or there exists $e''$ such that $e''\in \postset{(\postset{e})}$
  and $e''\le e'$.
\item In the second pass, we compute for each event $e$ the set
  $\conf{e}:=\{\,e'\mid e\conflict e'\,\}$, i.e., the set of events
  with which $e$ is in conflict. Here, we exploit that the conflict
  relation is inherited by causal successors: $e\conflict e'$ iff
  $\preset{e}\cap\preset{e'}\ne\emptyset$ or there exists $f$,$f'$ such
  that $f\le e$, $f'\le e'$, and $\preset{f}\cap\preset{f'}\ne\emptyset$.
  We traverse $E$ in topological order; each event~$e$ inherits
  the conflicts of its (direct) causal predecessors and obtains new conflicts
  with the set $\post{e'}$ for all events $e'$ with which it directly
  competes for some condition.
\item In the third pass, we finally compute a set $\rev{e}$ for each event
  $e$ such that $\rev{e}:=\{\,e'\mid e\reveals e'\,\}$. Here, we mainly
  exploit two facts: $e$ cannot reveal any events with which it is in conflict,
  and it reveals all events revealed by its causal predecessors:
  if $e''\reveals e'$ and $e''<e$, then $e\reveals e'$. We thus traverse
  $E$ in topological order; at each event, all known conflicts are discarded,
  and events from direct causal predecessors inherited. This leaves some
  events $e'$ for which the status is unknown (concurrent events and causal
  successors), and for these we check directly
  whether $\conf{e}\supseteq\conf{e'}$ (compare Lemma~\ref{lemma:witness}).
\end{enumerate}

\begin{algorithm}[!ht]
\caption{Computing the reveals relation}
\label{alg:reveals}
\small
\begin{algorithmic}
\STATE $\post{e}:=\{e\}$; \ $\conf{e}:=\emptyset$; \ $\rev{e}:=\{e\}$%
\ for all $e\in E$
\FORALL{$e\in E$ in inverse $\prec$-order}
\FORALL{$e'\in\postset{\postset{e}}$}
\STATE $\post{e}:=\post{e}\cup\post{e'}$
\ENDFOR
\ENDFOR
\FORALL{$e\in E$ in $\prec$-order}
\FORALL{$e'\in\preset{\preset{e}}$}
\STATE $\conf{e}:=\conf{e}\cup\conf{e'}$
\ENDFOR
\FORALL{$e'$ s.t.\ $\preset{e}\cap\preset{e'}\ne\emptyset$}
\STATE $\conf{e}:=\conf{e}\cup\post{e'}$
\ENDFOR
\ENDFOR
\FORALL{$e\in E$ in $\prec$-order}
\FORALL{$e'\in\preset{\preset{e}}$}
\STATE $\rev{e}:=\rev{e}\cup\rev{e'}$
\ENDFOR
\STATE $E':=E\setminus(\rev{e}\cup\conf{e})$;
\FORALL{$e'\in E'$}
\IF {$\rev{e}\supseteq\rev{e'}$}
\STATE $\rev{e}:=\rev{e}\cup\{e'\}$
\ENDIF
\ENDFOR
\ENDFOR
\end{algorithmic}
\end{algorithm}

Figure~\ref{alg:reveals} shows a version of the algorithm in pseudo-code.
Notice that if $\post{\cdot}$, $\conf{\cdot}$, and $\rev{\cdot}$ are
stored as bitsets (containing one bit for every event in $E$),
then almost all operations can be implemented using basic logical
operations on bitsets. In the first two passes, the number of such
operations is bounded by the number of arcs in $U$. In the third
pass, the number of operations is bounded by the pairs $(e,e')$ such
that $e'\notin(\rev{e}\cup\conf{e})$, that is by $|E|^2$ in the worst case.
However, it turns out that in most cases the number of such checks is
comparatively small.

\subsection{Computing \emph{reveals} for a single pair}
\label{sec:pair}

We briefly discuss the question of how to decide $x\reveals y$ for a
single pair of events $x,y$. If one is interested in individual pairs,
such a procedure may well be more efficient than the one from
Section~\ref{sec:fullrel} because it allows to limit the events one has
to consider.

Assume that $x,y$ are events of some unfolding $U$, of which at least
the prefix $\cone{x}\cup\cone{y}$ is known. (We assume that neither
$x\#y$ nor $x>y$ hold, otherwise the solution is trivial.) Denote by
$\#_\mu[y]:=\{\,z\mid z\in\#[y] \land \forall z':(z'<z\to z'\notin\#[y]\,\}$
the set of $<$-minimal conflicts of $y$, its so-called \emph{root conflicts}.
Due to results from~\cite{Haar-tac10} we know that $x\reveals y$ iff
$\#[x]\supseteq \#_\mu[y]$. To find a witness, it suffices therefore to
find an event $z$ that is not in conflict with $x$, but
a root conflict of $y$; the latter implies that
$\preset{z}\cap\precone{y}\ne\emptyset$.

We propose the following: First, mark the conditions in
$\precone{y}$ as `goals'. Secondly, mark all conditions and places
in conflict with $x$ as `useless' (they cannot produce a witness), as well
as all elements of $\precone{x}$ (which can equally not produce a witness
by assumption). One then regards the remaining non-`useless' events up to
the height given by Lemma \ref{le:U2}, either by unfolding them on-the-fly
or by following them on a pre-computed prefix. A witness is found if one such
`non-useless' events consumes a `goal' condition.

\section{Experiments}
\label{sec:exp}

We implemented the theoretical and algorithmical results of 
the preceding sections and evaluated them experimentally.
The problems we wanted to address were the following:
\begin{itemize}
\item What is the value of $K$ (as given by Lemma~\ref{le:U2}) for
	medium-sized nets?
\item Provided a prefix is available, how efficiently can one determine
  $\reveals$, using Algorithm~\ref{alg:reveals}?
\end{itemize}

As inputs, we chose the safe Petri net examples supplied 
by the PEP tool~\cite{grahlmann1997pep}.
Table~\ref{tab:stats} provides some statistics on the nets
we used, such as the number of places and transitions, as well as the
bound $K$ according to Lemma~\ref{le:U2} for each particular net.
We obtained $K$ by modifying the Mole unfolding tool~\cite{mole}. Normally,
Mole is used to compute finite complete prefixes; for our experiments,
we modified its cutoff criterion so that it would compute the unfolding
prefix $U_2$. We also give the time, in seconds, to compute the said prefix
in the rightmost column.

\begin{table}[!ht]
\caption{Net statistics and computation of $K$}
\centering
\tabcolsep0.68mm
\rowcolors[]{3}{gray!20}{white}
\small
{
\begin{tabularx}{5.6cm}{lrrrr}
\toprule
\rowcolor{white}
Petri net & $|P|$ & $|T|$ & $K$ & Time/s \\
\midrule
buf100             & 200 & 101 & 201 & 2.1 \\
elevator           &  59 &  74 &  80 & 0.3 \\
gas\_station       &  30 &  18 &  18 & 0.1 \\
mutual             &  62 &  67 &  -- & t/o \\
parrow             &  77 &  54 &  91 & 1.6 \\
peterson           &  27 &  31 &  34 & 0.1 \\
reader\_writer\_2  &  53 &  60 &  29 & 2.3 \\
sdl\_arq\_deadlock & 202 & 183 &  37 & 0.1 \\
sdl\_arq           & 208 & 234 & 129 & 0.2 \\
sdl\_example       & 323 & 471 &  71 & 0.1 \\
sem                &  26 &  25 &  35 & 0.1 \\
\bottomrule
\end{tabularx}
}
\label{tab:stats}

\end{table}
              
To make the experiments more interesting, we excluded non-cyclic examples,
where $K$ would be obvious. For the rest, the computation of $K$ succeeded
except in one case (\texttt{mutual}, more than 10 minutes). To give some
indications, the size of a \emph{complete} prefix in these cases
was between several dozen and a few thousand events, whereas the size of $U_2$
was between several hundred and several ten thousands of events.
By contrast, the computation of $K$ failed for
another set of larger benchmarks provided by Mole, whose complete prefixes
already have a size of 10,000 and more events.

To answer the second question, we implemented Algorithm~\ref{alg:reveals}
in Java. Our program took a pre-computed prefix and computed the
relation $\reveals$
on it, using the \texttt{BitSet} class for most operations.
The results are summarized in Table~\ref{tab:unf2}. As one can see, the
algorithm works well even for several tens of thousands of events, usually
computing the relation in a matter of seconds.

We detail the time for the three passes of the algorithm (all
times are in seconds); in almost each case, we have the same
ordering of computation times. The computation of the causal relation
($\mathit{post}$) takes hardly significant time, the second pass
for the computation of the conflict relation ($\mathit{conf}$)
takes a little more time, and the third pass for the computation of the
reveals relation ($\mathit{rev}$) slightly dominates the computation
time.

\begin{table}[!ht]
\caption{Running times of Algorithm~\ref{alg:reveals}}
\centering
\tabcolsep0.7mm
\rowcolors[]{3}{gray!20}{white}
\small
{
\begin{tabularx}{7.5cm}{lrrrrr}
\toprule
\rowcolor{white} 
Petri net & Events & post & conf & rev \\
\rowcolor{white} 
  & & (Time/s) & (Time/s) & (Time/s) \\
\midrule
bds\_1.sync&12900&0.13&0.19&0.30\\
buf100&17700&0.17&0.12&0.25\\
byzagr4\_1b&14724&0.18&0.19&0.68\\
dpd\_7.sync&10457&0.11&0.15&0.24\\
dph\_7.dlmcs&37272&0.56&0.91&2.10\\
elevator75&234879&15.84&22.58&97.47\\
elevator&5586&0.05&0.05&0.13\\
elevator\_4&16856&0.17&0.27&0.38\\
fifo20&100696&2.92&3.72&22.88\\
ftp\_1.sync&83889&2.08&3.61&6.78\\
furnace\_3&25394&0.29&0.47&0.95\\
gas\_station&2861&0.01&0.01&0.01\\
key\_4.fsa&67954&1.40&2.19&4.62\\
parrow&85869&2.47&4.17&9.51\\
peterson&72829&1.60&2.54&5.23\\
q\_1.sync&10722&0.11&0.15&0.30\\
q\_1&7469&0.08&0.09&0.17\\
reader\_writer\_2&20229&0.24&0.37&0.53\\
rw\_12.sync&98361&2.36&5.14&6.36\\
rw\_12&49179&0.68&1.25&1.70\\
rw\_1w3r&15401&0.15&0.22&0.50\\
rw\_2w1r&9241&0.10&0.11&0.25\\
sdl\_arq&2691&0.03&0.03&0.09\\
sem&19689&0.20&0.23&0.61\\
\bottomrule
\end{tabularx}
}
\label{tab:unf2}
\end{table}

\section{Conclusion}
\label{sec:conc}

We presented theoretical and algorithmic contributions towards the computation
of the \emph{reveals} relation. 
The analysis in \cite{Haar-tac10} had only
provided the proof that
$a \reveals b$ could be decided on \emph{some} bounded prefix of the unfolding;
but the bound (see Theorem \ref{th:oldresult}) was prohibitively large, and an
efficient procedure for computing $\reveals$ was lacking. The present paper
closes this theoretical and practical gap. Our results show that with a suitable
cutoff-criterion, the complete finite prefix $U_2$ is sufficient to obtain the
$\reveals$-relation on $U_1$. Moreover, an efficient algorithm for computing
$\reveals$ on finite occurrence nets has been proposed and tested;
the experimental results clearly show that $\reveals$ can be obtained and used
in practice.

The theory of reveals can be further developed in the lines of  \cite{BalChaHaa11}, where 
a dedicated logic (called ERL) is introduced for expressing generalized reveals
relation of the form "if all events from set A occur, then at least one event
from set B must eventually occur", and the problem 
of synthesizing occurrence nets from ERL formulas is solved. 
The study of further variants of logics for concurrency in the light 
of the recent results has only just begun.

In addition, we intend to extend \emph{reveals}-based analysis to other
Petri net classes such as Time nets and contextual nets, and to
exploit it in applications that include diagnosis and testing.

\bibliographystyle{eptcs}
\bibliography{GandALF2011}

\end{document}